\documentclass[a4paper,11pt]{article}
\usepackage{pos}
\newcommand{\sh}[1]{#1\hskip -6pt  / }

\title{Non-nucleonic degrees of freedom and the spin structure of the deuteron}

\author{Misak Sargsian }

\affiliation{Physics Department, Florida International University\\
  Miami, FL, 33199 USA}

\emailAdd{sargsian@fiu.edu}

\abstract{Electro-disintegration of the deuteron at large $Q^2$ currently represents on of the most promising 
reactions which allows to probe the bound  nuclear state at internal momenta comparable to the rest mass of the nucleon.
Large internal momentum in this case makes  non-nuncleonic states energetically more feasible and the question 
that we address is what  are the signatures that will indicate the existence of such states in the ground state of the 
nuclear wave function. 
 To probe such states we developed a light-front formalism for relativistic description of a composite 
 pseudo-vector system in which  emerging proton and  neutron are observed  in electro-disintegration reaction.
In leading high energy approximation our calculations show the  possibility of 
the existence of a new 
``incomplete" P-state-like  structure in the deuteron at extremely large internal momenta.
The incompleteness of the  observed P-state violates  the 
angular condition  for the momentum distribution, which can happen only if the deuteron contains non-nucleonic structures,
such as $\Delta\Delta$, $N^*N$ or hidden color components. 
Because such states have distinctive angular momentum ($l=1$) they significantly modify the polarization properties of 
the deuteron wave function.  As a result in addition to angular anisotropy of 
the LF momentum  distribution of the nucleon in the deuteron one predicts strong modification of 
the tensor polarization asymmetry of the deuteron  beyond the  S- and D- wave predictions at large 
internal momenta in the deuteron. }
\FullConference{25th International Spin Physics Symposium (SPIN 2023)\\
 24-29 September 2023\\
 Durham, NC, USA\\}

\begin{document}
\maketitle

\section{Introduction}
Understanding  the dynamics of 
the transition between hadronic to quark-gluon phases of matter is one of the outstanding issues of strong 
interaction physics. 
For cold dense nuclear matter such transitions are relevant for  the dynamics of 
superdense nuclear matter that can exist  at the cores of neutron stars and can set 
the limits for the  matter density before it collapses to  a black hole.
There are few  options to investigate such transitions. For example,
studying nuclear medium modification of quark-gluon structure of  bound nucleons by probing EMC effects
 \cite{EuropeanMuon:1983wih,Frankfurt:1988nt,Hen:2016kwk} especially in semi-inclusive processes which
 allow to control inter-nucleon distances\cite{Melnitchouk:1996vp}.  The other, rather different example is to
 study   implications of the transition of baryonic to quark matter in the cores of neutron stars by searching large mass  
($\approx 2.08 M_{\odot}$) neutron stars\cite{Fonseca:2021wxt}  with radii  $R< 10$\cite{Miller:2021qha}.

In the present work, new approach\cite{Sargsian:2022rmq} is suggested in probing 
baryon-quark transition by exploring deuteron at extremely large internal momenta.

\section{Deuteron on the light front~(LF)}
Non-relativistic picture of the deuteron suggests that 
the observations of total isospin, $I=0$,  total spin, $J=1$  and positive parity, $P$,  together with the  relation,
$P= (-1)^l$,  indicate  that the deuteron consists of bound proton and neutron in S- and D- partial wave states.

However, for the deuteron structure with internal momenta comparable to  the nucleon rest mass 
the nonrelativistic framework is not valid requiring a consistent account for the relativistic effects.
There are  several theoretical approaches for  accounting  relativistic effects in the deuteron wave function 
(see e.g.  Refs.\cite{Frankfurt:1977vc,Buck:1979ff,Arnold:1980zj,Dymarz:1986km,Carbonell:1995yi} and the 
reviews\cite{Frankfurt:1981mk,Miller:2000kv,Garcon:2001sz,Gilman:2001yh,Carbonell:1998rj}). In our approach
the  relativistic effects are accounted for similar to the one used in QCD (see e.g. \cite{Feynman:1973xc,Brodsky:1997de})
for  calculation of  quark distribution  in  hadrons, in which light-front (LF) description of 
the scattering process allows to suppress vacuum fluctuations that overshadow  the composite structure of the hadron. 
 In this approach  one needs to identify the process in which 
the deuteron structure is probed. For this  we consider high-momentum transfer electrodisintegration process:
\vspace{-0.2cm}
\begin{equation}
e + d \rightarrow e^\prime + p + n,
\label{reaction}
\vspace{-0.2cm}
\end{equation}
in which one of the nucleons are struck by the incoming probe and the spectator nucleon is probed with momenta comparable 
to the nucleon mass.  If one can neglect (or remove) the effects related to final state interactions of two outgoing nucleons, 
then the above reaction at high $Q^2$, measures the probability of observing  a proton and neutron in the deuteron 
with large relative  momenta.  
In such a formulation the deuteron is not a composite system consisting of a proton and neutron, but it is 
a composite pseudo - vector ($J=1$, $P=+$)  ``particle" from which one extracts a proton and neutron.
Thus we formulate the question not as how to describe relativistic motion of proton and neutron in the deuteron,  but
 how such a  proton and neutron are produced at   such   extreme conditions relating it  
to the dynamical  structure of the LF deuteron wave function. 
In such formulation the latter  may include internal elastic $pn \rightarrow pn$ as well as inelastic 
$\Delta\Delta\to pn$, $N^*N\to pn$ or $N_cN_c\to pn$ transitions. Here, $\Delta$ and $N^*$ denote $\Delta$-isobar and $N^*$ resonances,  while 
$N_c$ is a  color octet baryonic state contributing to  the hidden-color component in the deuteron.\\
The framework for calculation of reaction (\ref{reaction}) in the  relativistic domain  is the LF
approach (e.g. Ref.\cite{Frankfurt:1981mk,Miller:2000kv,Brodsky:1997de}) in which  one introduces 
the LF deuteron wave function:
\begin{equation}
\psi_{d}^{\lambda_d}(\alpha_i,p_\perp,\lambda_1\lambda_2) = - {\bar u(p_2,\lambda_2)\bar u(p_1,\lambda_1) \Gamma^\mu_{d} \chi_\mu^{\lambda_d}\over 
 {1 \over 2} ( m_d^2 - 4 {m_N^2 + p_\perp^2\over \alpha_i(2-\alpha_i)})\sqrt{2(2\pi)^3}} = 
 = -\sum\limits_{\lambda_1^\prime}\bar u(p_1,\lambda_1) 
\Gamma^\mu_{d}\gamma_5 {\epsilon_{\lambda_1,\lambda_1^\prime}\over \sqrt{2}}u(p_1,\lambda^\prime_1),
\vspace{-0.3cm}
 \label{dwave_lf}
\end{equation}
where $\alpha_i = 2{p_{i+}\over p_{d+}}$, ($i=1,2$) are LF momentum fractions of proton and neutron, outgoing from 
the deuteron with $\alpha_1+\alpha_2 = 2$ and in the second part we absorbed the 
propagator into the vertex function and used crossing symmetry.
Here  $u(p,\lambda)$'s are the LF bi-spinors of the proton and neutron\cite{Lepage:1980fj} and 
$\epsilon_{i,j}$ is the two dimensional Levi-Civita tensor, with $i,j=\pm 1$ nucleon helicity. Since the deuteron is a pseudo-vector 
``particle",  due to $\gamma_5$ in Eq.(\ref{dwave_lf}), the vertex $\Gamma^\mu_d$ is a four-vector which we can construct
in a general form that explicitly satisfies time reversal, parity and charge conjugate symmetries. Noticing  that  
at the   $ d\to pn$ vertex  on the light-front 
the  "-"  ($p^- = E-p_z$) components 
of the four-momenta of the particles are not conserved, in addition to the four-momenta of two nucleons, $p_1^\mu$ and $p_2^\nu$,  
one  has an additional four-momentum:
\begin{equation}
\Delta^\mu \equiv p_1^\mu + p_2^\mu - p_d^\mu   \equiv (\Delta^-, \Delta^+, \Delta_{\perp}) = (\Delta^-, 0, 0),
\label{DeltaDefLF}
\end{equation}
where
\vspace{-0.4cm}
\begin{eqnarray}
\vspace{-0.4cm}
& & \hspace{-0.8cm} \Delta^- =  p_1^- + p_2^- - p_d^- 
= 
{4\over p_d^+}\left[m_N^2 - {M_d^2\over 4} + k^2\right] ; k = \sqrt{{m_N^2 + k_\perp^2\over \alpha_1(2-\alpha_1)} - m_N^2} ;  \ \ \ \alpha_1 = {E_k + k_z\over E_k},
\vspace{-0.6cm}
\label{Delta-}
\end{eqnarray}
with  $E_k = {m^2 + k^2}$.
With  $p_1^\mu$, $p_2^\mu$ and $\Delta^\mu$ 4-vectors the $\Gamma_d^\mu$  is constructed in the form:
\vspace{-0.2cm}
\begin{eqnarray}
 & & \hspace{-0.2cm} \Gamma_d^{\mu}= \Gamma_{1}  \gamma^{\mu} +\Gamma_{2} 
 {{(p_1-p_2)^{\mu}}\over {2m_N}} + \Gamma_{3} 
 {{\Delta^{\mu}}\over {2m_N}}+
 \Gamma_{4} 
 {{(p_1-p_2)^{\mu} \sh{\Delta}}\over { 4m_N^2}}     \nonumber \\
& &    +  i \Gamma_{5} \frac{1}{4 m_N^{3}} 
\gamma_{5} \epsilon^{\mu \nu \rho \gamma}(p_{d})_\nu (p_1-p_2)_{\rho} (\Delta)_\gamma 
 +   \Gamma_{6} \frac{ \Delta^{\mu} \sh{\Delta}}{4m_N^{2}},
\label{vertex}
\end{eqnarray}
where $\Gamma_i$,($i=1,6$) are  scalar functions (see also Refs.\cite{Carbonell:1995yi,Carbonell:1998rj}). 

\section{High energy approximation} 
For the large $Q^2$ limit, the LF momenta for reaction (\ref{reaction})  are chosen as follows:
 \vspace{-0.2cm}
 \begin{eqnarray}
&&\hspace{-0.2cm}p_d^\mu  \equiv  (p_{d}^-, p_{d}^+ ,p_{d\perp}) =  \left({Q^2\over x\sqrt{s}}\left[1 + {x\over \tau} - \sqrt{1+ {x^2\over \tau}}\right] \right., 
 \left.                          {Q^2\over x\sqrt{s}} \left[1 + {x\over \tau} +  \sqrt{1+ {x^2\over \tau}}\right], 0_\perp\right)    \nonumber \\
&&q^\mu \equiv  (q^{-},q^{+}, q_{\perp}) =  \left({Q^2\over x\sqrt{s}}\left[1 - x + \sqrt{1+ {x^2\over \tau}}\right], \right.
\left.         {Q^2\over x\sqrt{s}}\left[1 - x - \sqrt{1+ {x^2\over \tau}}\right], 0_\perp\right),  
\label{refframeQ}
\end{eqnarray}
where $s = (q+p_d)^2$, $\tau={Q^2\over M_d^2}$ and $x = {Q^2\over M_dq_0}$, with $q_0$ being the virtual photon energy in 
the deuteron rest frame.  The high energy nature of this process results in,  $p_d^+ \sim \sqrt{Q^2}\gg m_N$, which makes 
$\Delta^-$ term to be suppressed by the large $p_d^+$ factor in  Eq.(\ref{Delta-}). As such we treat ${\Delta^-\over 2m_N}$  factor 
as a small parameter in the problem.

Analyzing now the vertex function (\ref{vertex}) 
one observes that $\Gamma_1$ and $\Gamma_2$ terms are explicitly leading order in  ${\cal O}^0({\Delta^-\over 2m_N})$.
The $\Gamma_3$ and $\Gamma_4$ terms enter with    order  ${\cal O}^1({\Delta^-\over 2m_N})$,
while the $\Gamma _6$ term enters as  ${\cal O}^2({\Delta^-\over 2m_N})$.  
The situation with the $\Gamma_5$ term is, however, different; since 
the covariant components:
$\Delta_+ = {1\over 2} \Delta^-$ and  $p_{d,-} = {1\over 2} p_d^+$,  
the  term with $\epsilon^{\mu + \perp -}$ is 
leading order (${\cal O}^0({\Delta^-\over 2m_N})$) due to the fact that the large $p_d^+$ factor is   canceled in the 
$p_{d,-} \Delta_+ = {1\over 4} p_d^+ \Delta^-$ combination.

Keeping the leading, ${\cal O}^0({\Delta^-\over 2m_N})$, terms in Eq.(\ref{vertex})  and using the boost invariance of the wave function 
we calculate it in the CM of the deuteron\cite{Sargsian:2022rmq} to  obtain:
 \vspace{-0.2cm}
\begin{eqnarray}
& & \hspace{-0.9cm} \psi_{d}^{\lambda_d}(\alpha_i,k_\perp)  \hspace{-0.1cm} =    -\hspace{-0.5cm} \sum\limits_{\lambda_2,\lambda_1,\lambda_1^\prime}
\hspace{-0.3cm}\bar u(-k,\lambda_2) 
 \hspace{-0.1cm}\left\{ \Gamma_1\gamma^\mu \hspace{-0.1cm}+ \hspace{-0.1cm}
\Gamma_2{{\tilde k}^\mu\over m_N}  + \right.    
\hspace{-0.3cm}  \left.    \sum\limits_{i=1}^{2}
 i\Gamma_5{1\over 8m^3_N}\epsilon^{\mu + i  -}p^{\prime +}_{d} k_{i} \Delta^{\prime-}\right\}   
  \gamma_5 \hspace{-0.1cm}{\epsilon_{\lambda_1,\lambda_i^\prime}\over \sqrt{2}} u(k,\lambda^\prime_1)s_\mu^{\lambda_d},   
\label{dwave_lf4}
\end{eqnarray}
where $\tilde k^\mu = (0,k_z,k_\perp)$  with $k_\perp = p_{1\perp}$,
$k^2 = k_z^2 + k^2_\perp$ and  $E_{k} = 
{\sqrt{S_{NN}}\over 2}$  and  $s_\mu^{\lambda_d} = (0,{\bf s^\lambda_d})$, with
$s_d^1 = - {1\over \sqrt{2}} (1,i,0)$,   $s_d^1 = {1\over \sqrt{2}} (1,-i,0)$,   $s_d^0 = (0,0,1)$
and
$p^{\prime +}_{d}   = \sqrt{s_{NN}}, \    
\Delta^{\prime-}  = {1\over \sqrt{s_{NN}}}\left[ {4(m_N^2 + k_\perp^2)\over\alpha_1(2-\alpha_1)}-M_d^2\right]$.
Since the term related to  
$\Gamma_5$ is proportional to  ${4(m_N^2 + k_\perp^2)\over\alpha_1(2-\alpha_1)}-M_d^2$, which 
diminishes at small momenta,  only the $\Gamma_1$ and $\Gamma_2$ terms will contribute in the  nonrelativistic limit  defining 
the  $S$- and $D$- components of the deuteron.  Thus, the LF wave function in Eq.(\ref{dwave_lf4}) provides a smooth transition to 
the non-relativistic deuteron wave function.
This can be seen by expressing Eq.(\ref{dwave_lf4})  through two-component spinors:
\vspace{-0.2cm}
\begin{eqnarray}
 \psi_d^{\lambda_d}(\alpha_1,k_t,\lambda_1,\lambda_2)  =  
\sum\limits_{\lambda_1^\prime}\phi^\dagger_{\lambda_2} \sqrt{E_k}\left[  {U(k)\over \sqrt{4\pi}} {\bf \sigma s_d^{\lambda_d}}\right. - 
  - \left.    {W(k)\over \sqrt{4\pi}  \sqrt{2}}\left( { 3{\bf (\sigma k)(k s_d^\lambda)}\over k^2} - {\bf \sigma s_d^\lambda} \right) +\right. & &    \nonumber \\  
  \left.    
 (-1)^{1+\lambda_d\over 2} P(k)Y_{1}^{\lambda_d}(\theta,\phi)\delta^{1,\mid \lambda_d\mid}
  \right]
{\epsilon_{\lambda_1,\lambda_1^\prime}\over \sqrt{2}} \phi_{\lambda^\prime_1}. & &
\label{WF_LF}
\end{eqnarray}
Here the first two terms have explicit $S$- and $D$- structures  where the radial functions are defined as: 
\begin{eqnarray}
& \hspace{-0.4cm} U(k)  = & {2\sqrt{4\pi} \sqrt{E_k}\over 3}\left[\Gamma_1(2+{m_N\over E_k}) + \Gamma_2{k^2\over m_N E_k}\right]\nonumber \\
& \hspace{-0.4cm} W(k)  = &  {2\sqrt{4\pi} \sqrt{2E_k}\over 3}\left[\Gamma_1(1-{m_N\over E_k}) -   \Gamma_2{k^2\over m_N E_k}\right].  
\label{radialwaves}
\end{eqnarray}
This relation is  known for   $pn$-component deuteron wave function\cite{Frankfurt:1981mk,Carbonell:1995yi}, which allows us to 
model the LF wave function through known radial $S$- and $D$- wave functions    evaluated at LF relative momentum $k$ defined  in
Eq.(\ref{Delta-}).

The new result is that due to the $\Gamma_5$ term  there is an additional leading contribution, 
which because of the relation  $Y^{\pm}_1(\theta,\phi) = \mp i\sqrt{3\over 4\pi}\sum\limits_{i=1}^{2}{ (k\times s_d^{\pm 1})_z\over k}$, has a 
$P$-wave like structure, where the $P$- radial function  is defined as:
\begin{eqnarray}
&  \hspace{-0.4cm}P(k)  = &  \sqrt{4\pi}  {\Gamma_5(k) \sqrt{E_k}\over \sqrt{3}}{k^3\over m_N^3}.   
\label{Pradialwave}
\end{eqnarray}
Note  that this term  is purely relativistic in origin: as it follows from Eq.(\ref{Pradialwave})  it has  an extra ${k^2\over m_N^2}$ factor 
in addition to the ${k^{l=1}\over m_N}$ term that  characterizes  the radial $P$-wave. 
Thus our result does not affect the known non-relativistic wave function.
 
The unusual feature of our  result is that the $P$-wave is ``incomplete", that is it  contributes only 
for $\lambda_d = \pm 1$ polarizations of the deuteron.
 
\section{Light front density matrix of the deuteron}
 One defines the unpolarized deuteron LF momentum distribution $n_d(k,k_\perp)$ and
density matrix \cite{Frankfurt:1981mk,Frankfurt:1988nt} as follows:
\vspace{-0.2cm}
\begin{equation}
\vspace{-0.2cm}
n_d(k,k_\perp)    =   {1\over 3}\sum\limits_{\lambda_d=-1}^{1}\mid \psi_d^{\lambda_d}(\alpha,k_\perp)\mid^2  \ \ \ 
\mbox{and} \ \ \ \rho_d({\alpha,k_\perp}) = {n_d(k,k_\perp)\over 2-\alpha}.
\label{momdist1}
\end{equation}
Using Eq.(\ref{WF_LF}) the  LF momentum distribution is expressed through the radial wave functions as follows:
\vspace{-0.2cm}
\begin{equation}
\vspace{-0.2cm}
n_d(k,k_\perp)    =   {1\over 3}\sum\limits_{\lambda_d=-1}^{1}\mid \psi_d^{\lambda_d}(\alpha,k_\perp)\mid^2 = 
=  {1\over 4\pi} \left( U(k)^2 +  W(k)^2 + {k_\perp^2\over k^2} P^2(k)\right)
\label{momdist2}
\end{equation}
with 
$\int \rho_d({\alpha,k_\perp}) {d\alpha\over \alpha} = 1$, $\int\alpha \rho_d({\alpha,k_\perp}) {d\alpha\over \alpha} = 1$ and 
$\int \left(U(k)^2 +  W(k)^2 + {2\over 3} P^2(k)\right)k^2 dk = 1$.
Due to the incompleteness of the $P$-wave structure our result predicts that LF momentum distribution for unpolarized 
deuteron  depends  explicitly on the transverse component of the relative momentum on the light front. This is highly unusual 
result, implication of which will be discussed in the next section.

For polarized deuteron the quantity that can be probed in the reaction~(\ref{reaction}) the tensor asymmetry which we define as:
\begin{equation}
A_T = {n_d^{\lambda_d = 1}(k,k_\perp) +  n_d^{\lambda_d = -1}(k,k_\perp) - 2  n_d^{\lambda_d = 0}(k,k_\perp)\over  n_d(k, k_\perp)}.
\label{t20}
\end{equation}
Here because of the same incompleteness of the $"P-wave"$ structure one may expect more sensitivity that for unpolarized momentum 
distribution.

\section{The new term and the non-nucleonic components in the deuteron:}
One of our main predictions is that the LF momentum distribution, Eq.(\ref{momdist2}) will explicitly depend on 
the transverse component of  the  deuteron internal momentum on the light front. 
Such  a dependence  is impossible for non-relativistic quantum mechanics of the 
deuteron since in this case the potential of the interaction is real (no inelasticities) and the solution of Lippmann-Schwinger equation 
for partial S- and D-waves satisfies the ``angular condition", according to which the momentum distribution in the unpolarized deuteron depends 
on the magnitude of the relative momentum only.  As we mentioned earlier, our result does not contradict  the properties of  non-relativistic deuteron wave function since, according to Eq.(\ref{Pradialwave}) the P-wave is purely relativistic in nature. 

In the relativistic 
domain the definition of the interaction  potential is not straightforward to allow  the use of quantum-mechanical arguments
in claiming  that the momentum distribution in Eq.(\ref{momdist2}) should satisfy the angular condition also in the relativistic case
(i.e. to be dependent only on the  magnitude of $k$). 

To check the situation in relativistic case one considers Weinberg type 
equation\cite{Weinberg:1966jm} on the light-front for NN scattering amplitudes, in which only nucleonic degrees are considered,
in the CM of the NN system.  One obtains\cite{Frois:1991wc}:
\begin{eqnarray}
&&T_{NN}(\alpha_i,k_{i\perp},\alpha_f,k_{f,\perp}) \equiv T_{NN}(k_{i,z},k_{i\perp},k_{f,z},k_{f,\perp}) =  V(k_{i,z},k_{i\perp},k_{f,z},k_{f,\perp}) \nonumber \\
& &  + \int V(k_{i,z},k_{i\perp},k_{m,z},k_{m,\perp}) 
 {d^3 k_m\over (2\pi)^3 \sqrt{m^2 + k_m^2}}{T_{NN}(k_{m,z},k_{m\perp},k_{f,z},k_{f,\perp})\over 4(k_m^2 - k_f^2)},
\label{TNN}
\end{eqnarray} 
where ``i", ``m" and ``f" subscripts correspond to initial, intermediate and final $NN$ states, respectively, and momenta $k_{i,m,f}$ 
are defined similar to Eq.(\ref{Delta-}). 

The realization of the angular condition for the relativistic case will require that the light-front potential in general to 
satisfy  the condition:
\begin{equation}
V(k_{i,z},k_{i\perp},k_{m,z},k_{m,\perp}) = V(\vec k_i^2, (\vec k_m-\vec k_i)^2).
\label{Vangcond}
\end{equation}
Such a  conditions  for the on-shell limit follows from   the Lorentz invariance of the $T_{NN}$ amplitude:
\begin{equation}
T_{NN}^{on \ shell}(k_{i,z},k_{i\perp},k_{m,z},k_{m,\perp}) = T^{on \ shell}_{NN}(\vec k_i^2, (\vec k_m-\vec k_i)^2)
\label{Tangcond}
\end{equation}
and the existence of the Born term in Eq.(\ref{TNN}) indicates that  the potential $V$ satisfies 
the same condition  in the  on-shell limit.   

For the off-shell potential, 
it was shown\cite{Frankfurt:1988nt,FSMS90,Frois:1991wc} 
that requirements for  the potential $V$ to satisfy
angular condition in the on-shell limit and 
that it  can be constructed through the series of elastic $pn$ scatterings  result
in a potential which is an analytic function of angular momentum.
With the assumption that the potential, analytically continued to the complex angular momentum space, does not diverge 
exponentially,  it was shown that the $V$ and $T_{NN}$ functions satisfy the angular condition (Eqs.(\ref{Vangcond},\ref{Tangcond})) in general.
Using the same potential to calculate the LF deuteron wave function will result in a momentum distribution 
dependent only  on 
the magnitude of the relative $pn$ momentum. This observation
requires a  consideration of  the   $pn$ component only in the deuteron.  

Inclusion of the inelastic transitions will completely change the LF equation for the $pn$ scattering. 
For  example, the contribution of  $N^*N$ transition to the elastic $NN$ scattering:
\vspace{-0.3cm}
\begin{eqnarray}
&&\hspace{-0.4cm}T_{NN}(k_{i,z},k_{i\perp},k_{f,z},k_{f,\perp}) =  \int V_{NN^*}(k_{i,z},k_{i\perp},k_{m,z},k_{m,\perp})  \nonumber \\
& & 
\times{d^3 k_m\over (2\pi)^3 \sqrt{m^2 + k_m^2}} {T_{N^*N}(k_{m,z},k_{m\perp},k_{f,z},k_{f,\perp})\over 4(k_m^2 - k_f^2+m^2_{N*} - m_N^2)},
\label{TNN*}
\end{eqnarray} 
will not require the condition of Eq.(\ref{Vangcond}) with the transition potential having also an imaginary component.  Eq.(\ref{TNN*})
can not be described with any combination of elastic $NN$ interaction potentials that satisfies the angular condition.
The same will be true also  for $\Delta\Delta\to NN$  and   $N_c,N_c\to NN$ transitions.

Thus one concludes that if the $\Gamma_5$ term is not zero  and results in a $k_\perp$ dependence of LF momentum distribution 
then it should originate from a non-nucleonic component in the deuteron.

\section{Predictions and estimate of the possible effects}
Our calculations predict three new effects, that in probing deuteron structure at very large internal momenta ($\ge m_N$) 
in reaction (\ref{reaction}):\\
\noindent 
- the LF momentum distribution should be enhanced compared to $S$- and $D$- wave contributions only;  \\
\noindent - there should be  angular anisotropy  in the LF momentum distribution; \\
\noindent - the tensor asymmetry should be significantly different as expected from $S$- and $D$- wave contributions only. 

Observation of all  the above effects will indicate a presence of non-nucleonic components in the deuteron wave function at 
large internal momenta.

\begin{figure}[h]
\begin{center}
\includegraphics[width=8.2cm,height=5.2cm]{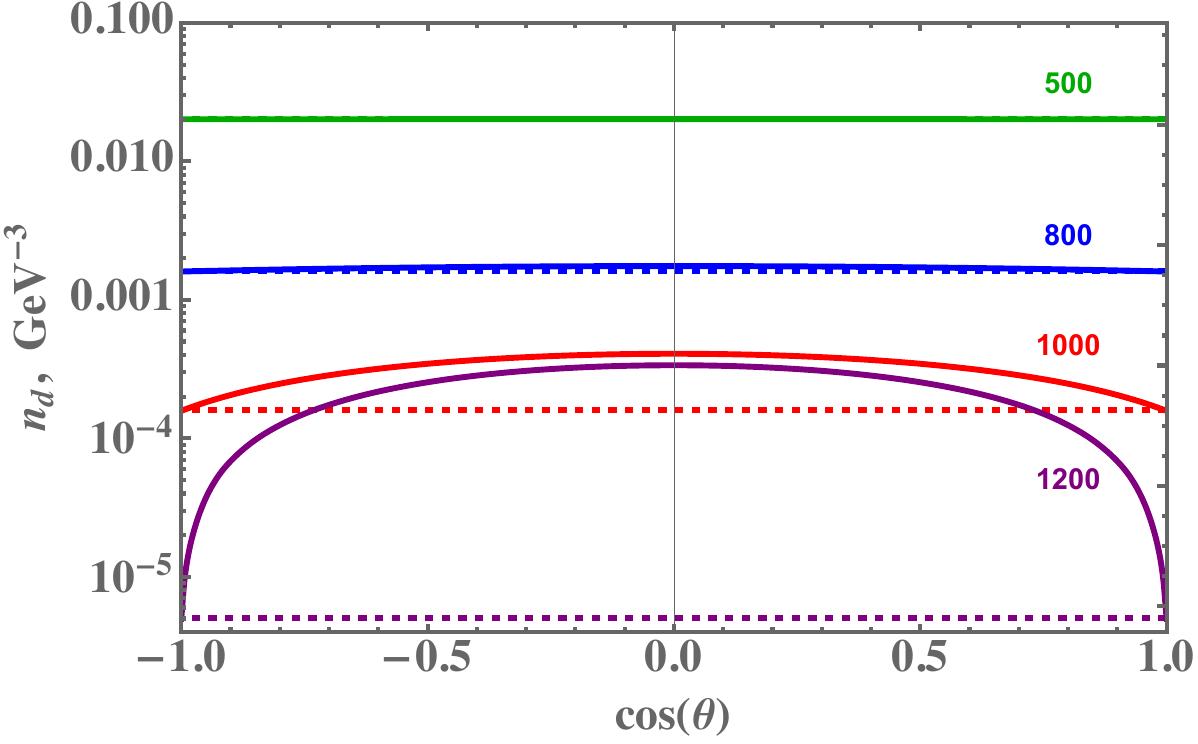}
\vspace{-0.5cm}
\end{center}
\caption{LF momentum distribution of the deuteron as a function of $\cos{\theta}$, for different values of $k$.
 Dashed lines - deuteron with $pn$ component only, solid lines - with $P$-wave like component included.}
\label{mbfigures1}
\end{figure}

To give quantitative estimates of the possible effects we evaluate the $\Gamma_5$ vertex function assuming  two color-octet baryon transition to 
the $pn$ system ($N_cN_c\to pn$) through the one-gluon exchange,  parameterizing it in the dipole form ${A\over  (1 + {k^2\over 0.71})^2}$. The parameter $A$ is estimated by assuming 1\% contribution to the total normalization from the $P$ wave. The latter is consistent 
with the experimental estimation in Ref.\cite{deltadelta}  of 0.7\%. In Fig.\ref{mbfigures1} we consider the dependence of the momentum 
distribution of Eq.(\ref{momdist2}) as a function of $\cos{\theta} = {(\alpha-1)E_k\over k}$ for different values of $k$. Notice that if the momentum distribution is 
generated by the $pn$ component only, the angular condition is satisfied, and  no dependence should be observed.

As the figure shows one may expect  measurable angular dependence  at $k\gtrsim 1$~GeV/c, which is consistent with the expectation that the 
inelastic transition in the deuteron corresponding to the non-nucleonic  components takes place at $k\gtrsim 800$~MeV/c.  

\begin{figure}[h]
\begin{center}
\includegraphics[scale=0.36]{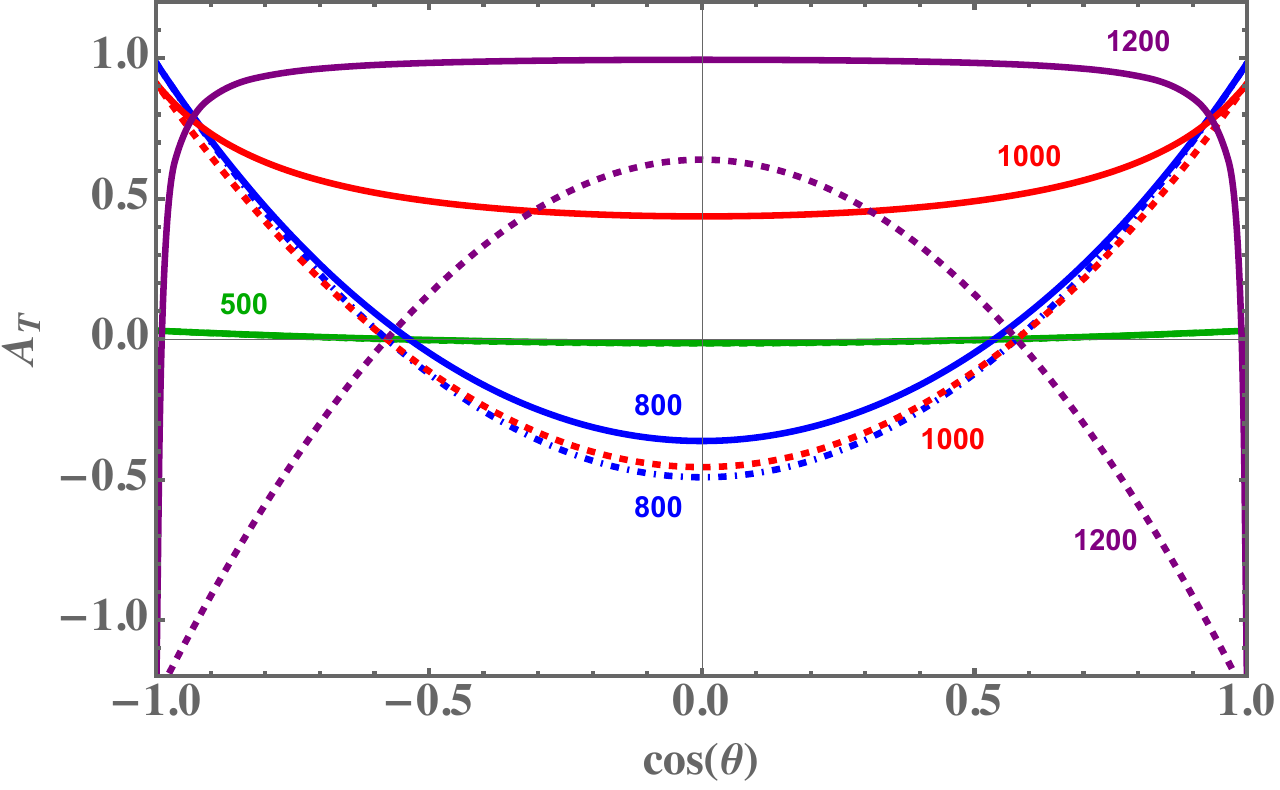}
\vspace{-0.5cm}
\end{center}
\caption{Tensor asymmetry as a function of  $\cos{\theta}$ for different $k$.  Dashed lines - deuteron with $pn$ component only,
solid lines - with $P$ component included.}
\label{mbfigures2}
\end{figure}

For tensor polarized deuteron we estimated the effect using Eq.~(\ref{t20}). The results are presented in Fig.\ref{mbfigures2}.
As the figure shows the presence of a non-nucleonic component will 
be visible   already at $k\approx 800$~MeV/c, resulting in a qualitative difference  
in the asymmetry at larger momenta as compared with the asymmetry predicted  by the deuteron wave function with 
a $pn$-component only.
 
The reason why small, 1\% effect in overall normalization  gives a  large 
measurable effect  in LF momentum distribution and asymmetry at $k\ge 1$GeV/c  is due to the fact that the observed
``incomplete P-wave" structure  enters with the  ${p^2\over m_N^2}$ prefactor (see Eq.(\ref{Pradialwave})),   which 
significantly amplifies the effect at very large internal momenta.

\section{Outlook on experimental verification of the predicted  effects} 
The predictions discussed in the previous section which are related to the existence of  non-nucleonic component in the deuteron 
wave function  can be be verified at CM momenta $k \gtrsim 1$~GeV/c.  These seem 
an incredibly large momenta to be measured in experiment. However, the first such measurement at high $Q^2$ disintegration of the 
deuteron  has already been performed at Jefferson Lab\cite{HallC:2020kdm} reaching $k\sim 1$~GeV/c.  It is intriguing that the 
results of this  measurement qualitatively disagree with  predictions based on conventional deuteron wave functions once $k\gtrsim 800$~MeV/c.
Moreover the data seems to indicate the enhancement of momentum distribution as predicted in our calculations.
  
The planned new measurements \cite{Boeglin:2014aca} will significantly improve the quality of the data allowing possible verification of the
second prediction, that is  the existence of angular asymmetry for LF momentum distribution.

What concerns to the tensor asymmetry, currently there is a significant efforts in measuring  high $Q^2$ deuteron electro-disintegration 
processes at Jefferson Lab employing polarized deuteron target\cite{Slifer:2013vma}.  The possibility of  performing such measurements 
will significantly improve the validity of any observation that will suggest the existence of non-nucleonic component in the deuteron 
wave function at very large internal momenta.

 It is worth mentioning that the analysis of exclusive  deuteron disintegration  experiments will require a careful account for competing nuclear 
effects such as final state  interactions, (FSI) for which there has been significant  theoretical and experimental progress during the last decade\cite{Frankfurt:1996xx,Sargsian:2001ax,Sargsian:2009hf,Boeglin:2015cha}.
The advantage of high energy scattering is that the eikonal regime is established which makes FSI to be strongly isolated in 
transverse kinematics and be  suppressed in near colinear directions. Additionally the comparison with the first high $Q^2$ experimental 
data\cite{Boeglin:2015cha} indicates that the accuracy of FSI calculations increases with $Q^2$ which will allow a meaningful analysis of new 
high $Q^2$ data.

If the experiments will  not find the discussed signatures of non-nucleonic components  then they will set a new limit on 
the dominance of the $pn$ component at instantaneous high nuclear densities that corresponds to $\sim 1$~GeV/c internal momentum 
in the deuteron.  However if predictions are confirmed,  they  will motivate theoretical modeling of non-nucleonic components in 
the deuteron, such as $\Delta\Delta$, $N^*N$ or hidden-color $N_cN_c$ that can reproduce the observed results. 
In both cases the results of such 
studies will  advance the understanding of the dynamics of high density nuclear matter and the relevance of the quark-hadron transitions.

 {\bf Acknowledgments:}
This work is supported by the U.S. DOE Office of Nuclear Physics grant DE-FG02-01ER41172.

\end{document}